\documentclass{Interspeech2024}

\interspeechcameraready 

\title{Lightweight Audio Segmentation for Long-form Speech Translation\thanks{* Equal contribution.}}

\name[affiliation={1*}]{Jaesong}{Lee}
\name[affiliation={1,2*}]{Soyoon}{Kim}
\name[affiliation={1}]{Hanbyul}{Kim}
\name[affiliation={2}]{Joon Son}{Chung}

\address{
  $^1$NAVER Cloud, South Korea\\
  $^2$KAIST, South Korea
}
\email{\{jaesong.lee, soyoon.kim, hanbyul.kim\}@navercorp.com, joonson@kaist.ac.kr}


\newcommand{\MaxSegLen}{\texttt{maxlen}}
\newcommand{\MinSegLen}{\texttt{minlen}}

\newcommand{\Period}{\texttt{$\langle$.$\rangle$}}
\newcommand{\Question}{\texttt{$\langle$?$\rangle$}}
\newcommand{\Comma}{\texttt{$\langle$,$\rangle$}}

\newcommand{\FairseqST}{Fairseq-ST}

\begin{document}

\maketitle

\begin{abstract}
Speech segmentation is an essential part of speech translation (ST) systems in real-world scenarios. Since most ST models are designed to process speech segments, long-form audio must be partitioned into shorter segments before translation.
Recently, data-driven approaches for the speech segmentation task have been developed.
Although these approaches improve overall translation quality, a performance gap exists due to a mismatch between the models and ST systems.
In addition, the prior works require large self-supervised speech models, which consume significant computational resources.
In this work, we propose a segmentation model that achieves better speech translation quality with a small model size.
We propose an ASR-with-punctuation task as an effective pre-training strategy for the segmentation model.
We also show that proper integration of the speech segmentation model into the underlying ST system is critical to improve overall translation quality at inference time.
\end{abstract}

\section{Introduction}

Speech translation (ST), which converts speech signals from one language into text in another language, helps facilitate communication between people who speak different languages and helps overcome language barriers.
Integrating an automatic speech recognition (ASR) component with a machine translation (MT) component is commonly referred to as a cascaded architecture, and it is the traditional and common approach for the ST task~\cite{ney1999speech}.

Recently, there has been a growing interest in end-to-end (E2E) methods that directly translate spoken source language to target language text using a single sequence-to-sequence model~\cite{berard2016listen,weiss17_interspeech}.
Since E2E ST doesn't produce intermediate speech recognition results, it can prevent ASR errors from propagating to the translation model.
It can also improve latency and model size because it combines the ASR and MT modules into a single model for inference.
However, this approach is still less accurate than the cascade system~\cite{sperber2020speech,agrawal2023findings}.

Although both cascade and E2E ST systems have been actively developed, the models are designed to process {segmented} speech due to constraints on model architecture and training conditions.
Long-form speech must be segmented in advance to use the ST system in real-world scenarios where segmentation is not available.
However, until recently, it has been under-explored how the segmentation impacts the overall quality of the ST system.

If a segmentation method is not \emph{matched} to the underlying ST system, it could lead to low-quality translation results~\cite{salesky2023evaluating}.
In~\cite{wicks2022sentence}, two common failure modes due to the mismatch are discussed.
When a segment is too long or contains multiple sentences, the translation may omit some of the input, called a deletion error.
On the other hand, if a segment is too short or does not contain a proper sentence, the translation may contain phrases not in the input, referred to as an insertion error or {hallucination}~\cite{raunak2021curious,lee2018hallucinations}.
Thus, it is essential to produce segments of appropriate duration and with a single complete sentence to meet the requirements of the underlying ST system.

Several segmentation methods for ST have been previously introduced in the literature ~\cite{potapczyk2020srpols,gaido2021beyond,gallego2021end,radford2023whisper}.
Pause-based segmentation using voice activity detection (VAD) is commonly employed as a preliminary step for ST systems~\cite{potapczyk2020srpols,gaido2021beyond,gallego2021end}.
Another widely used strategy involves length-based segmentation techniques, where speech is divided into segments according to heuristic principles~\cite{gallego2021end,radford2023whisper}. 
For cascaded speech translation systems, there are works on re-segmentation of ASR output text~\cite{cho2012segmentation,wan2021segmenting}.
Also, it is proposed to interpret predictions of ASR and ST models for fixed-size chunks as segmentation~\cite{yoshimura2020end,huang2023e2e,polak2023long,shu2023cif}.

Recently, data-driven approaches for audio segmentation have been proposed~\cite{shas,fukuda2022speech,fukuda2024improving}, which consist of a neural network encoder and predict segmentation at frame level.
The methods have been shown to improve segmentation performance compared to the traditional methods.
However, the translation quality of the proposed methods is still behind the one of oracle segmentation, as a mismatch exists between the two segmentation results~\cite{salesky2023evaluating}.
Also, the models are usually based on large self-supervised models, such as wav2vec 2.0~\cite{baevski2020wav2vec},
whose computational cost would be a hurdle for deploying ST system on mobile devices.

In this paper, we aim to improve the end-to-end segmentation modeling for long-form speech translation while significantly reducing the number of model parameters.
We propose that the ASR-with-punctuation task~\cite{nozaki2022interspeech,kim23_interspeech}, the joint task of speech recognition and punctuation prediction, is an effective pre-training task for the segmentation model.
In addition, we show that tuning the segmentation model at inference time is essential to the overall translation quality, and provide an analysis of the mismatch between the segmentation model and ST system.

Our contributions are as follows:
\begin{itemize}
    \item We propose a pre-training strategy for the segmentation model using the ASR-with-punctuation task and show that the proposed pre-training strategy improves the segmentation accuracy and the final translation quality.
    \item We show that reducing the mismatch between the segmentation model and ST systems is crucial, due to varying characteristics among ST systems.
    \item We demonstrate that the proposed segmentation model achieves better translation quality than the prior methods, and its size is 8\% to 14\% smaller than that of the previous works.
\end{itemize}

\section{Architecture}
\label{sec:architecture}

We formulate the audio segmentation task as a frame-level classification task, following previous works~\cite{shas,fukuda2022speech,fukuda2024improving}.
The segmentation model is trained to predict a frame-level label sequence for a fixed-length audio input.
The label sequence $(l_1, \cdots, l_T)$ consists of the binary label $l_t \in \{0, 1\}$, which indicates whether the $t$-th frame is a part of segment ($l_t = 1$) or not ($l_t = 0$).

The model consists of an encoder layer and a linear layer.
First, the model converts a sequence of acoustic features to a sequence of output vectors $(\mathbf{e}_1, \cdots, \mathbf{e}_T)$,
where $\mathbf{e}_t \in \mathbb{R}^D$ represents the $t$-th output vector.
Conformer-M~\cite{gulati2020conformer} architecture is used for the encoder layer, which accepts log-mel acoustic features and gives the output sequence with a 40ms stride.

The output vector $\mathbf{e}_t$ transformed to a probability $p_t$ by a linear layer. The value $p_t$ indicates the probability that $t$-th frame is a part of the segment, and it is computed as:
\begin{equation*}
    p_t = \mathsf{sigmoid}(\mathsf{Linear}_{D \rightarrow 1}(\mathbf{e}_t)).
\end{equation*}
During training, the cross-entropy loss is used to match $p_t$ to the segmentation label $l_t$.
The length of the input audio is 20 seconds following prior works~\cite{shas,fukuda2024improving}.

Note that the model size is much smaller than the those of previous works.
SHAS~\cite{shas} has 201M parameters and SHAS-FTPT~\cite{fukuda2024improving} has 349M parameters, as they are based on XLS-R~\cite{babu2021xlsr}, a large self-supervised wav2vec 2.0~\cite{baevski2020wav2vec} model.
On the other hand, our model has 27.3M parameters, which are only \textbf{14\%} of SHAS and \textbf{8\%} of SHAS-FTPT.
Consequently, the model is more suitable for lightweight applications including streaming and on-device scenarios.

\subsection{Inference}
\label{sec:inference}

At inference time, the long-form audio is partitioned into fixed-size chunks of 20 seconds, with 2-second of overlap.
For overlapped frames, the two probability values are averaged.
Then, the output probability sequence is processed into a list of segments.

For the processing, SHAS~\cite{shas} introduces pDAC (probabilistic Divide-And-Conquer), which recursively splits a large segment into smaller segments.
pDAC has two hyper-parameters {\MinSegLen} and {\MaxSegLen} so that the resulting segment is always longer than {\MinSegLen} and shorter than {\MaxSegLen}.
pDAC has a drawback in that it often produces segments longer than oracle segments~\cite{fukuda2024improving} because it tends not to split a segment shorter than \MaxSegLen.

SHAS-FTPT~\cite{fukuda2024improving} proposes pTHR, a threshold-based algorithm, which also ensures the length of the segment is between {\MinSegLen} and {\MaxSegLen}.
pTHR has a drawback in that if the predicted segment is longer than {\MaxSegLen}, it is split into fixed-size segments of {\MaxSegLen}.

To this end, we use a simple algorithm as follows:
\begin{itemize}
    \item The predicted probability $p_t$ is converted to a binary label $l_t = \mathbb{I}[p_t > 0.5]$ and consecutive positive labels form a segment.
    \item Segments shorter than {\MinSegLen} are discarded.
    \item Segments longer than {\MaxSegLen} are split at $\hat{t}$-th position, where $\hat{t} = \mathsf{argmin}_t(p_t)$.
    \item Following the previous methods~\cite{shas,fukuda2024improving}, we expanded each segment by 0.06 seconds.
\end{itemize}
We found that {\MaxSegLen} is an important hyper-parameter that should be tuned for integrating the segmentation model and underlying ST system.
See Section~\ref{sec:integration} for discussion and Section~\ref{sec:experiments} for experiments.

\section{Pre-training via ASR-with-punctuation}
\label{sec:pretraining}

\begin{figure}[t]
  \centering
  \includegraphics[width=1.0\linewidth]{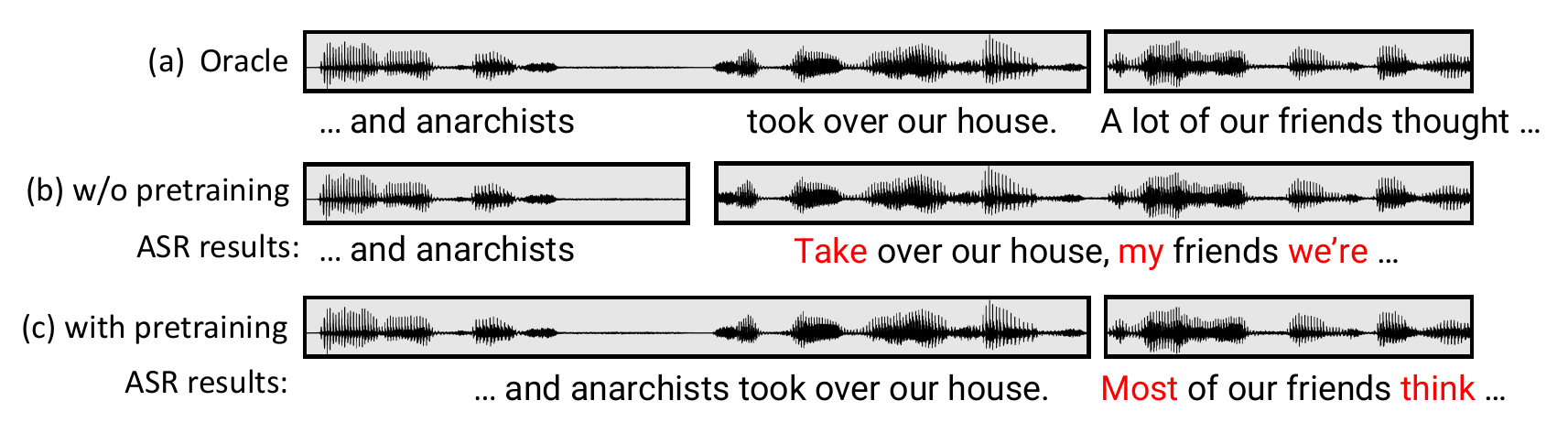}
  \vspace{-0.5cm}
  \caption{(a) Oracle segmentation and its corresponding reference text.
  (b) prediction of segmentation model \textbf{without} pre-training, and its corresponding ASR results.
  (c) prediction of segmentation model \textbf{with} pre-training, and its corresponding ASR results.
  ASR errors are colored red.
  See Section~\ref{sec:pretraining} for details.
  }
  \label{fig:pretrain_sentence_bounary}
  \vspace{-0.3cm}
\end{figure}

The ASR-with-punctuation task aims to recognize text and predict punctuation at the same time~\cite{mimura2021end,nozaki2022interspeech,kim23_interspeech,radford2023whisper}.
In particular, recent works~\cite{nozaki2022interspeech,kim23_interspeech} showed that ASR models based on Connectionist Temporal Classification (CTC)~\cite{graves2006connectionist} are suitable for ASR-with-punctuation with high accuracy.

CTC has a characteristic that the model predicts a text label for each frame (including a special label \texttt{<blank>}, which indicates no label corresponds to the frame), and the position of the predictive label is well-aligned to the corresponding speech utterance~\cite{kurzinger2020ctc}.
Thus, the behavior of the CTC model is closely related to the segmentation task.

We argue that both the ASR-with-punctuation task and the sentence-level segmentation task require understanding the grammar of sentences to certain degree.
Some punctuation marks, including period {\Period} and question mark {\Question}, indicate the end of the sentence.
To predict them, the CTC model needs to understand where the sentence boundaries are.
Also, the frame that predicts such marks is likely to be the position in which the segment of the sentence ends.

On the other hand, a spoken sentence may contain long pauses between utterances. In this case, the CTC model should not emit end-of-sentence marks during pauses, and the segmentation model should not partition the sentence at pauses as well.

Therefore, we expect that the CTC model with punctuation prediction learns features related to sentence structure, which are also helpful for sentence-level segmentation.
To this end, we propose to pre-train the encoder of the segmentation model with punctuation CTC task.

Following~\cite{kim23_interspeech}, we concatenate two segments of the ASR corpus at training time. This prevents the ASR model from predicting period symbol {\Period} at the end regardless of input, as the symbol is placed at the end of the sentence in many ASR corpus.
We apply Intermediate CTC~\cite{interctc} following previous works on ASR-with-punctuation~\cite{nozaki2022interspeech,kim23_interspeech}.

Figure~\ref{fig:pretrain_sentence_bounary} shows two segmentation results with their corresponding ASR transcriptions, one from the model without pre-training and the other with pre-training.
Without pre-training, the model relies on pauses for segmentation, causing mis-segmentation around the phrase ``took over our house''.
This also leads to a critical mis-transcription (``\emph{Take} over our house''), as the phrase is not a proper sentence, while the ASR model is likely to be trained with complete sentences.
More importantly, incomplete sentences from the mis-segmentation are more prone to mis-translation due to missing information~\cite{wicks2022sentence}.

On the other hand, the pre-trained model does not rely only on long pauses.
It successfully predicts the sentence boundary, which also leads to fewer transcription errors in the ASR model.
This illustrates the segmentation ability based on the speech content, not only the acoustic statistics.
We show experimental results that pre-training improves the overall translation quality in Section~\ref{sec:experiments}.

\section{Integration to speech translation system}
\label{sec:integration}

\begin{figure}[t]
  \centering
  \includegraphics[width=1.0\linewidth]{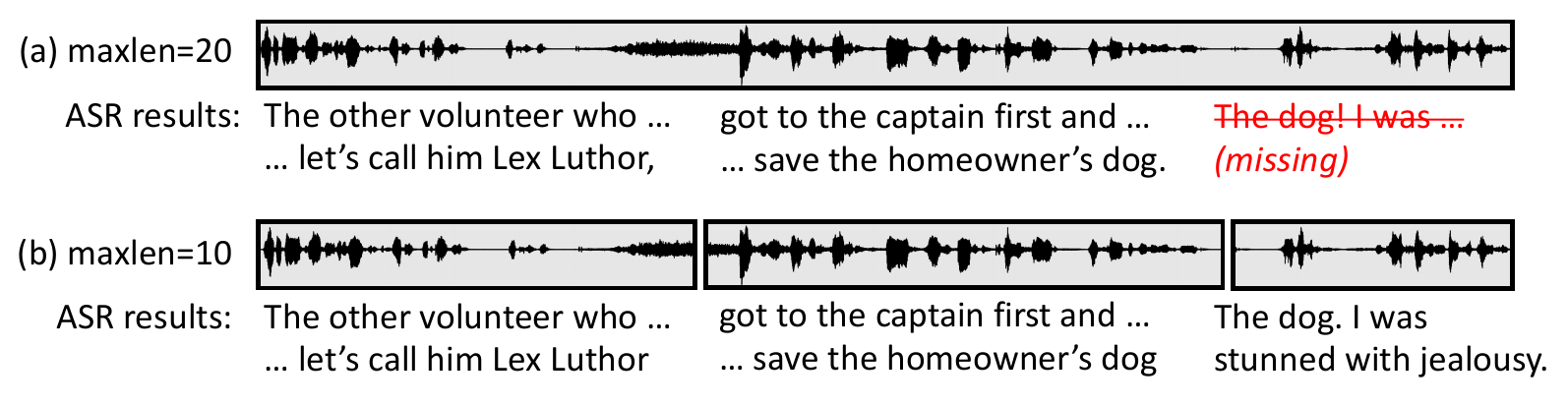}
  \vspace{-0.5cm}
  \caption{Segmentation and corresponding ASR results with two different {\MaxSegLen} configurations. Note that the two results are inferred from the same segmentation model.
  See Section~\ref{sec:integration} for details.
  }
  \vspace{-0.5cm}
  \label{fig:maxlen20_deletion}
\end{figure}

Due to the variety in model architectures and training conditions, ST systems require specific conditions of input speech segment for high-quality translation.
Importantly, if the audio segmentation is \emph{mismatched} to the segmentation used to train the ST system, its translation quality may be significantly degraded~\cite{wicks2022sentence,salesky2023evaluating}.

There are two well-known failure cases due to the mismatch.
If the speech segment input is too long or contains too many sentences, the ST model may fail to translate and drop a significant part of the input, causing \emph{deletion} errors~\cite{wicks2022sentence}.
On the other hand, if the segment is too short, it may not contain a complete sentence, which leads to significant \emph{addition} errors, also called as \emph{hallucination}~\cite{lee2018hallucinations,raunak2021curious,wicks2022sentence}.
Also, the exact notation of sentence boundary varies over translation corpora and target domains, contributing to the mismatch problem~\cite{wicks2022sentence,minixhofer2023wheres,matusov2019customizing,tsiamas2023segaugment}.
Therefore, to prevent such failure modes, it is essential to {match} the segmentation model and ST model so that the segmentation model produces speech segments that the ST model can handle well.

At the inference time, there is a hyper-parameter {\MaxSegLen} regarding the maximum length of the speech segment, as described in Section~\ref{sec:inference}.
The hyper-parameter can be tuned to reduce the mismatch between the segmentation model and the ST system.
If the segment produced by the segmentation model is longer than {\MaxSegLen}, it is partitioned into smaller segments.
We found that the partitioned segments tend to contain near-complete sentences rather than incomplete phrases.
This is because the pre-training task proposed in Section~\ref{sec:pretraining} improves the understanding of sentence-level boundaries and prevents non-linguistic splits at long pauses.

Figure~\ref{fig:maxlen20_deletion} shows an example of segmentation configuration and corresponding ASR results, where the ASR model in use tends to require short segments for better recognition accuracy.
The two results are inferred from the same segmentation model, except that the first setting~(a) uses {\MaxSegLen} 20 and the second setting~(b) uses {\MaxSegLen} 10.
The setting~(a) yields a long segment that matches the oracle segmentation.
However, the ASR model is not able to handle such long input, causing deletion errors towards the end of the segment.

On the other hand, the setting~(b) forces the segmentation model to yield short segments.
As a result, the audio is split into three segments, and the third segment still contains a complete sentence.
The first two segments contain incomplete phrases, as it is impossible to split the sentence.
Nonetheless, setting~(b) gives a better translation overall, as the last sentence is correctly recognized.

Note that the example is specific to the ASR model in use -- when the other ST model is used, the ST model successfully produces high-quality translation results for the long segment.
We show experimental results with various ST systems in Section~\ref{sec:experiments}.


\section{Experiments}
\label{sec:experiments}

MuST-C~\cite{mustc} is a multilingual speech corpus that can be used for automatic speech recognition (ASR), speech translation (ST), and audio segmentation tasks.
It consists of long-form speech of English TED Talks, sentence-level segmentation labels, transcription (with punctuation) and translation for each segment.

For the segmentation task, we use two language pairs of MuST-C v2, English-German (En-De) and English-Japanese (En-Ja).
For the pre-training task in Section~\ref{sec:pretraining}, we use the English transcription of En-De pairs.

We evaluate the segmentation models on the En-De and En-Ja ST tasks.
For the En-De ST task, we employ two ST systems. The first is the {\FairseqST}\footnote{\url{https://github.com/facebookresearch/fairseq/tree/main/examples/speech_text_joint_to_text}, En-De MuST-C model}~\cite{fairseq} E2E ST model, which has also been used in prior works~\cite{shas,fukuda2024improving}.
The second is SeamlessM4T-v2\footnote{\url{https://github.com/facebookresearch/seamless_communication}, seamlessM4T\_v2\_large model}~\cite{barrault2023seamless}. The model supports English ASR, En-De MT, and En-De E2E ST.
We employ the ASR and MT models for the cascaded ST system, as its translation quality is significantly better than the one of the E2E ST model, and it is possible to measure the accuracy of the ASR task as well as the ST task.

For the En-Ja ST task, we employ SeamlessM4T-v2 for the ASR model and employ two En-Ja MT models, SeamlessM4T-v2 and JParaCrawl\footnote{En-Ja big model}~\cite{morishita2020jparacrawl}, for cascaded ST.

For evaluating the ST system for long-form speech, mwerSegmenter~\cite{matusov2005evaluating} is used to re-align the results of ASR and ST models to the reference text. Then, sacreBLEU~\cite{post2018call} is used to measure BLEU scores~\cite{papineni2002bleu}.
For cascaded ST, we also measure word error rates (WERs) of the ASR model.

We compare our segmentation model to SHAS\footnote{\url{https://github.com/mt-upc/SHAS}}~\cite{shas} (En-De and En-Ja) and SHAS-FTPT\footnote{\url{https://github.com/ahclab/Wav2VecSegmenter}}~\cite{fukuda2024improving} (En-De) using the model parameters released by the authors.

For the choice of {\MaxSegLen}, we evaluate the segmentation model for two ST systems using 8, 10, 15, 20, and 30 seconds and report the best result for each ST system.
For SHAS and SHAS-FTPT, we also evaluate the models with {\MaxSegLen} reported in the papers.
For {\MinSegLen}, we use 0.2 seconds following SHAS and SHAS-FTPT.

\subsection{Results}
\label{sec:main_results}

\begin{table}[t]
\centering
\caption{Results of MuST-C En-De speech translation.
}
\vspace{-0.3cm}
\label{table:en_de}
\begin{tabular}{@{}l@{\hspace{0.5em}}c@{\hspace{0.5em}}c c @{\hspace{0.5em}} c @{\hspace{0.5em}} c@{}}
\toprule
&  & \texttt{max} & {\FairseqST} & \multicolumn{2}{c}{SeamlessM4T-v2} \\
Segmentation & \#param & \texttt{len} & BLEU$\uparrow$ & WER\%$\downarrow$ & BLEU$\uparrow$  \\
\midrule
Oracle &       &    & {26.90} & {14.58} & {29.49} \\
\midrule
SHAS & 201M  & 10 & 24.99 & \underline{14.70} & \underline{27.89} \\
& & 20 & \underline{25.57} & 23.61 & 25.09 \\
\midrule
SHAS-FTPT & 349M  & 15 & 25.95 & \underline{17.00} & \underline{28.13} \\
& & 28 & \underline{26.30} & {17.98} & {27.70} \\
\midrule
Proposed  & \textbf{27.3M} & 10 & 25.78 & \textbf{12.85} & \textbf{28.80} \\ 
& & 20 & \textbf{26.66} & 15.51 & 28.45 \\

\bottomrule
\end{tabular}
\end{table}

\begin{table}[t]
\centering
\caption{Results of MuST-C En-Ja speech translation.
``JParaCrawl'' indicates the cascade ST system of SesmlessM4T-v2 ASR and JParaCrawl MT models.
}
\vspace{-0.3cm}
\label{table:en_ja}
\begin{tabular}{@{}l@{\hspace{0.5em}}c@{\hspace{0.5em}}c c c @{\hspace{0.5em}} c@{}}
\toprule
&  & \texttt{max} & \multicolumn{2}{c}{SeamlessM4T-v2} & JParaCrawl\\
Segmentation & \#param & \texttt{len} & WER\%$\downarrow$ & BLEU$\uparrow$ & BLEU$\uparrow$  \\
\midrule
Oracle &       &    & {12.91} & {10.16} & {11.91} \\
\midrule
SHAS & 201M  & 8 & \underline{13.56} & \underline{9.03} & {11.34} \\
& & 18 & 21.90 & {8.44} & \underline{11.59} \\
\midrule
Proposed  & \textbf{27.3M} & 10 & \textbf{12.45} & \textbf{10.08} & 11.65 \\ 
& & 20 & 14.49 & 9.70 & \textbf{11.83} \\

\bottomrule
\end{tabular}
\end{table}

\begin{figure}[t]
  \centering
  \includegraphics[width=0.9\linewidth]{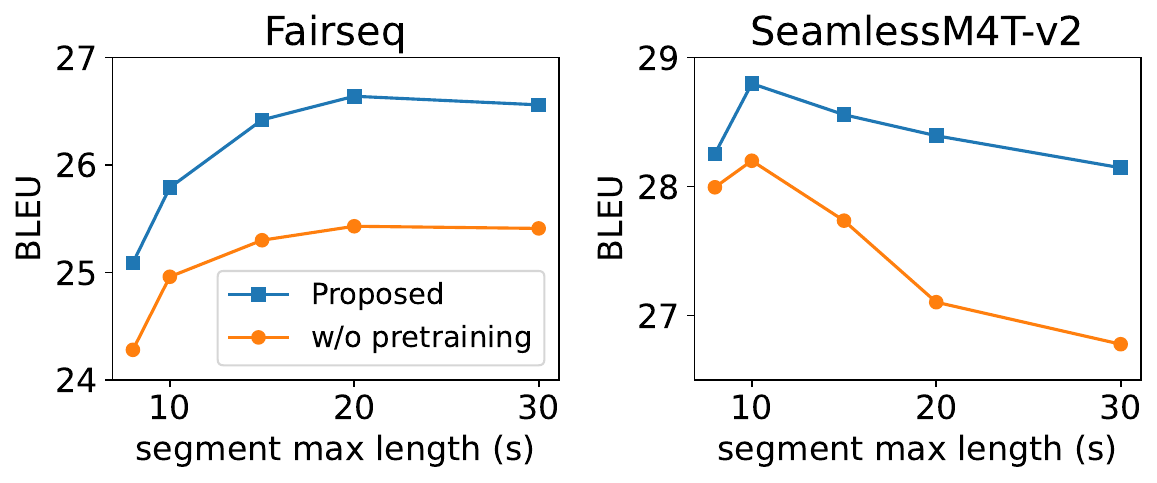}
  \vspace{-0.3cm}
  \caption{En-De BLEU scores for various {\MaxSegLen}.
  }
  \label{fig:length_bleu}
\end{figure}

Table~\ref{table:en_de} shows the evaluation results for the En-De task.
We found our model consistently outperforms the baseline models, whereas the number of parameters in our model (27.3M) is much smaller than the baseline models (201M and 349M).

We found {\FairseqST} and SeamlessM4T-v2 require different {\MaxSegLen} for the best accuracy -- $\MaxSegLen=20$ for Fairseq and $\MaxSegLen = 10$ for SeamlessM4T-v2.
Figure~\ref{fig:length_bleu} shows En-De BLEU scores for various {\MaxSegLen}.
It shows that the BLEU score decreases by 1.5 BLEU points if {\MaxSegLen} is not tuned properly.

Table~\ref{table:en_ja} shows the evaluation results for the En-Ja task.
Similar to En-De, our model performs consistently better than the baseline model,
and SeamlessM4T-v2 MT yields the best BLEU for shorter {\MaxSegLen} while JParaCrawl MT does for longer {\MaxSegLen}.

For the SeamlessM4T-v2 ASR task, we found shorter {\MaxSegLen} leads to lower WER.
Note that the WER of $\MaxSegLen = 10$ is even lower than the WER of oracle segmentation for both En-De and En-Ja.
This is because the oracle segmentation contains long segments, causing deletion errors on SeamlessM4T-v2.
When $\MaxSegLen = 8$, we obtained the lowest WERs of \textbf{12.35} for En-De and \textbf{11.70} for En-Ja.

However, for the following MT task, short {\MaxSegLen} and low WER do not necessarily improve the BLEU score.
We found the best BLEU scores are obtained from $\MaxSegLen=10$ for the SeamlessM4T-v2 MT model and $\MaxSegLen=20$ for the JParaCrawl MT model.
This could be explained by the fact that some non-critical ASR errors (e.g., ``went and met'' becomes ``went to meet'') do not necessarily cause fatal translation errors, and MT models generally require long input for better translation.

The overall results highlight that each ST system has different requirements for the translation quality, and for cascaded ST, it depends on the both the ASR and MT models.
Therefore, for reliable evaluation of segmentation models, employing multiple ST systems for measurement is important.

For the effectiveness of ASR pre-training in Section~\ref{sec:pretraining}, we measure BLEU scores of two segmentation models, with or without ASR pre-training. The result is shown in Figure~\ref{fig:length_bleu}.
We see the ASR pre-training consistently improves BLEU on the two ST systems.

\subsection{Evaluation of ASR punctuation prediction}
\label{sec:eval_punct_asr}

\begin{table}[t]
\centering
\caption{Punctuation F1 scores of SeamlessM4T-v2 ASR.
}
\vspace{-0.3cm}
\label{table:asr_punct_f1}
\begin{tabular}{lccccc}
\toprule
& & \multicolumn{4}{c}{Punctuation F1 $\uparrow$ (\%)} \\
Segmentation & \#param & avg. & {\Period} & {\Question} & {\Comma} \\
\midrule
Oracle & & 75.60 & 84.43 & 76.06 & 66.31 \\
\midrule
SHAS & 201M & 65.15 & 69.59 & 66.82 & 59.03 \\
SHAS-FTPT & 349M & 65.50 & 71.59 & 66.67 & 58.23 \\
\midrule
Proposed & \textbf{27.3M} & \textbf{67.79} & \textbf{73.62} & \textbf{69.91} & \textbf{59.84} \\
\bottomrule
\end{tabular}
\end{table}

For the ASR-with-punctuation task, wrong audio segmentation can lead to incorrect punctuation mark predictions, as the punctuation mark is difficult to predict correctly if the segment does not contain a complete sentence.
For example, Figure~\ref{fig:pretrain_sentence_bounary}~(a) shows that the ASR model replaces a period {\Period} with a comma {\Comma} for incorrect segmentation.

To this end, we measure F1 scores of three punctuation marks, period {\Period}, question mark {\Question}, and comma {\Comma} of SeamlessM4T-v2 ASR results from various segmentation models used in Section~\ref{sec:main_results}.
Following prior works on ASR-with-punctuation~\cite{nozaki2022interspeech,kim23_interspeech}, the output of the ASR model is aligned to the reference text for measurement.

Table~\ref{table:asr_punct_f1} shows the punctuation F1 scores of ASR results for the segmentation methods.
It shows that our model consistently outperforms baseline models for three punctuation marks, implying the ability to understand sentence structure. This, in turn, improves the ST performance.


\section{Conclusion}

We propose a lightweight end-to-end audio segmentation modeling for improving the quality of long-form speech translation.
We propose to use ASR-with-punctuation as a pre-training task for audio segmentation and show experimental improvements.
We emphasize the need for the match between the segmentation model and the speech translation system to achieve optimal translation quality, and show that tuning the inference-time hyper-parameter reduces the mismatch problem.
Furthermore, we show the proposed segmentation model achieves better translation quality with a model size of only 8\% to 14\% of the baseline models.

\clearpage
\bibliographystyle{IEEEtran}
\bibliography{mybib}

\begin{thebibliography}{10}
\providecommand{\url}[1]{#1}
\csname url@samestyle\endcsname
\providecommand{\newblock}{\relax}
\providecommand{\bibinfo}[2]{#2}
\providecommand{\BIBentrySTDinterwordspacing}{\spaceskip=0pt\relax}
\providecommand{\BIBentryALTinterwordstretchfactor}{4}
\providecommand{\BIBentryALTinterwordspacing}{\spaceskip=\fontdimen2\font plus
\BIBentryALTinterwordstretchfactor\fontdimen3\font minus \fontdimen4\font\relax}
\providecommand{\BIBforeignlanguage}[2]{{%
\expandafter\ifx\csname l@#1\endcsname\relax
\typeout{** WARNING: IEEEtran.bst: No hyphenation pattern has been}%
\typeout{** loaded for the language `#1'. Using the pattern for}%
\typeout{** the default language instead.}%
\else
\language=\csname l@#1\endcsname
\fi
#2}}
\providecommand{\BIBdecl}{\relax}
\BIBdecl

\bibitem{ney1999speech}
H.~Ney, ``Speech translation: Coupling of recognition and translation,'' in \emph{Proc. ICASSP}, 1999.

\bibitem{berard2016listen}
A.~Berard, O.~Pietquin, C.~Servan, and L.~Besacier, ``Listen and translate: A proof of concept for end-to-end speech-to-text translation,'' 2016.

\bibitem{weiss17_interspeech}
R.~J. Weiss \emph{et~al.}, ``{Sequence-to-Sequence Models Can Directly Translate Foreign Speech},'' in \emph{Proc. Interspeech}, 2017.

\bibitem{sperber2020speech}
M.~Sperber and M.~Paulik, ``Speech translation and the end-to-end promise: Taking stock of where we are,'' in \emph{Proc. ACL}, 2020.

\bibitem{agrawal2023findings}
M.~Agarwal \emph{et~al.}, ``Findings of the {IWSLT} 2023 evaluation campaign,'' in \emph{Proc. IWSLT}, 2023.

\bibitem{salesky2023evaluating}
E.~Salesky \emph{et~al.}, ``Evaluating multilingual speech translation under realistic conditions with resegmentation and terminology,'' in \emph{Proc. IWSLT}, 2023.

\bibitem{wicks2022sentence}
R.~Wicks and M.~Post, ``Does sentence segmentation matter for machine translation?'' in \emph{Proc. WMT}, 2022.

\bibitem{raunak2021curious}
V.~Raunak, A.~Menezes, and M.~Junczys-Dowmunt, ``The curious case of hallucinations in neural machine translation,'' in \emph{Proc. NAACL-HLT}, 2021.

\bibitem{lee2018hallucinations}
K.~Lee \emph{et~al.}, ``Hallucinations in neural machine translation,'' 2018.

\bibitem{potapczyk2020srpols}
T.~Potapczyk and P.~Przybysz, ``{SRPOL}{'}s system for the {IWSLT} 2020 end-to-end speech translation task,'' in \emph{Proc. IWSLT}, 2020.

\bibitem{gaido2021beyond}
M.~Gaido, M.~Negri, M.~Cettolo, and M.~Turchi, ``Beyond voice activity detection: Hybrid audio segmentation for direct speech translation,'' in \emph{Proc. ICNLSP}, 2021.

\bibitem{gallego2021end}
G.~I. G{\'a}llego \emph{et~al.}, ``End-to-end speech translation with pre-trained models and adapters: {UPC} at {IWSLT} 2021,'' in \emph{Pro. IWSLT}, 2021.

\bibitem{radford2023whisper}
A.~Radford \emph{et~al.}, ``Robust speech recognition via large-scale weak supervision,'' in \emph{Proc. ICML}, 2023.

\bibitem{cho2012segmentation}
E.~Cho, J.~Niehues, and A.~Waibel, ``Segmentation and punctuation prediction in speech language translation using a monolingual translation system,'' in \emph{Proc. IWSLT}, 2012.

\bibitem{wan2021segmenting}
D.~Wan \emph{et~al.}, ``Segmenting subtitles for correcting {ASR} segmentation errors,'' in \emph{Proc. EACL}, 2021.

\bibitem{yoshimura2020end}
T.~Yoshimura, T.~Hayashi, K.~Takeda, and S.~Watanabe, ``End-to-end automatic speech recognition integrated with {CTC}-based voice activity detection,'' in \emph{Proc. ICASSP}, 2020.

\bibitem{huang2023e2e}
W.~R. Huang \emph{et~al.}, ``{E2E} segmentation in a two-pass cascaded encoder {ASR} model,'' in \emph{Proc. ICASSP}, 2023.

\bibitem{polak2023long}
P.~Pol{\'a}k and O.~Bojar, ``Long-form end-to-end speech translation via latent alignment segmentation,'' 2023.

\bibitem{shu2023cif}
Y.~Shu \emph{et~al.}, ``A {CIF}-based speech segmentation method for streaming {E2E} {ASR},'' \emph{IEEE Signal Processing Letters}, 2023.

\bibitem{shas}
I.~Tsiamas, G.~I. Gállego, J.~A.~R. Fonollosa, and M.~R. Costa-jussà, ``{SHAS: Approaching optimal Segmentation for End-to-End Speech Translation},'' in \emph{Proc. Interspeech}, 2022.

\bibitem{fukuda2022speech}
R.~Fukuda, K.~Sudoh, and S.~Nakamura, ``{Speech Segmentation Optimization using Segmented Bilingual Speech Corpus for End-to-end Speech Translation},'' in \emph{Proc. Interspeech}, 2022.

\bibitem{fukuda2024improving}
------, ``Improving speech translation accuracy and time efficiency with fine-tuned wav2vec 2.0-based speech segmentation,'' \emph{TASLP}, 2024.

\bibitem{baevski2020wav2vec}
A.~Baevski, Y.~Zhou, A.~Mohamed, and M.~Auli, ``wav2vec 2.0: A framework for self-supervised learning of speech representations,'' in \emph{Proc. NeurIPS}, 2020.

\bibitem{nozaki2022interspeech}
J.~Nozaki, T.~Kawahara, K.~Ishizuka, and T.~Hashimoto, ``{End-to-end Speech-to-Punctuated-Text Recognition},'' in \emph{Proc. Interspeech}, 2022.

\bibitem{kim23_interspeech}
H.~Kim, S.~Seo, L.~Lee, and S.~Baek, ``{Improved Training for End-to-End Streaming Automatic Speech Recognition Model with Punctuation},'' in \emph{Proc. Interspeech}, 2023.

\bibitem{gulati2020conformer}
A.~Gulati \emph{et~al.}, ``Conformer: Convolution-augmented transformer for speech recognition,'' in \emph{Proc. Interspeech}, 2020.

\bibitem{babu2021xlsr}
A.~Babu \emph{et~al.}, ``{XLS-R}: Self-supervised cross-lingual speech representation learning at scale,'' 2021.

\bibitem{mimura2021end}
M.~Mimura, S.~Sakai, and T.~Kawahara, ``An end-to-end model from speech to clean transcript for parliamentary meetings,'' in \emph{Proc. APSIPA}, 2021.

\bibitem{graves2006connectionist}
A.~Graves, S.~Fern\'{a}ndez, F.~Gomez, and J.~Schmidhuber, ``Connectionist temporal classification: labelling unsegmented sequence data with recurrent neural networks,'' in \emph{Proc. ICML}, 2006.

\bibitem{kurzinger2020ctc}
L.~K{\"u}rzinger \emph{et~al.}, ``{CTC}-segmentation of large corpora for german end-to-end speech recognition,'' in \emph{International Conference on Speech and Computer}, 2020.

\bibitem{interctc}
J.~Lee and S.~Watanabe, ``Intermediate loss regularization for ctc-based speech recognition,'' in \emph{Proc. ICASSP}, 2021.

\bibitem{minixhofer2023wheres}
B.~Minixhofer \emph{et~al.}, ``Where{'}s the point? self-supervised multilingual punctuation-agnostic sentence segmentation,'' in \emph{Proc. ACL}, 2023.

\bibitem{matusov2019customizing}
E.~Matusov, P.~Wilken, and Y.~Georgakopoulou, ``Customizing neural machine translation for subtitling,'' in \emph{Proc. WMT}.

\bibitem{tsiamas2023segaugment}
I.~Tsiamas, J.~Fonollosa, and M.~Costa-juss{\`a}, ``{S}eg{A}ugment: Maximizing the utility of speech translation data with segmentation-based augmentations,'' in \emph{Proc. EMNLP 2023}, 2023.

\bibitem{mustc}
M.~A. Di~Gangi \emph{et~al.}, ``{M}u{ST}-{C}: a {M}ultilingual {S}peech {T}ranslation {C}orpus,'' in \emph{Proc. NAACL-HLT}, 2019.

\bibitem{fairseq}
M.~Ott \emph{et~al.}, ``fairseq: A fast, extensible toolkit for sequence modeling,'' in \emph{Proc. ACL (Demonstrations)}, 2019.

\bibitem{barrault2023seamless}
L.~Barrault \emph{et~al.}, ``Seamless: Multilingual expressive and streaming speech translation,'' 2023.

\bibitem{morishita2020jparacrawl}
M.~Morishita, J.~Suzuki, and M.~Nagata, ``{JP}ara{C}rawl: A large scale web-based {E}nglish-{J}apanese parallel corpus,'' in \emph{Proc. LREC}, 2020.

\bibitem{matusov2005evaluating}
E.~Matusov, G.~Leusch, O.~Bender, and H.~Ney, ``Evaluating machine translation output with automatic sentence segmentation,'' in \emph{Proc. IWSLT}, 2005.

\bibitem{post2018call}
M.~Post, ``A call for clarity in reporting {BLEU} scores,'' in \emph{Proc. WMT}, 2018.

\bibitem{papineni2002bleu}
K.~Papineni, S.~Roukos, T.~Ward, and W.-J. Zhu, ``{B}leu: a method for automatic evaluation of machine translation,'' in \emph{Proc. ACL}, 2002.

\end{thebibliography}

\end{document}